\documentclass[12pt,a4paper]{article}
\usepackage[utf8]{inputenc}
\usepackage[english]{babel}
\usepackage{amsmath,amsfonts,amssymb,amsthm}
\usepackage[left=2cm,right=2cm,top=2cm,bottom=2cm]{geometry}
\usepackage{placeins}
\usepackage[overload]{empheq}
\usepackage{tikz}
\usetikzlibrary{shapes,arrows}
\tikzstyle{decision} = [diamond, draw, fill=blue!20, 
    text width=4.5em, text badly centered, node distance=3cm, inner sep=0pt]
\tikzstyle{block} = [rectangle, draw, fill=blue!20, 
    text width=5em, text centered, rounded corners, minimum height=4em]
\tikzstyle{line} = [draw, -latex']
\tikzstyle{cloud} = [draw, ellipse,fill=red!20, node distance=3cm,
    minimum height=2em]
\allowdisplaybreaks
\begin{document}
\title{Continuous-time stochastic processes for the spread of COVID-19 disease simulated via a Monte Carlo approach and comparison with deterministic models}
\date{}
\author{Fabiana Calleri\thanks{Universit\`{a} degli Studi di Catania, Dipartimento di Matematica e Informatica, Viale Andrea Doria 6, 95125 Catania, Italy  ({\tt fabianacalleri@yahoo.it}).}
\and Giovanni Nastasi\thanks{Universit\`{a} degli Studi di Catania, Dipartimento di Matematica e Informatica, Viale Andrea Doria 6, 95125 Catania, Italy  ({\tt g.nastasi@unict.it}).}
\and Vittorio Romano\thanks{Universit\`{a} degli Studi di Catania, Dipartimento di Matematica e Informatica, Viale Andrea Doria 6, 95125 Catania, Italy  ({\tt romano@dmi.unict.it}).} }
\maketitle
\begin{abstract}
Two stochastic models are proposed to describe the evolution of the COVID-19  pandemic. In the first model the population is partitioned into four compartments: susceptible $S$,  infected $I$, removed $R$ and  dead people $D$. In order to have a cross validation, a deterministic version of such a model is also devised which is represented by a system of  ordinary differential equations with delays.
In the second stochastic model  two further compartments are added: the class $A$ of asymptomatic individuals and the class $L$ of isolated infected people. Effects such as social distancing measures are easily included and the consequences are analyzed. 

Numerical solutions are obtained with Monte Carlo simulations. Quantitative predictions are provided which can be useful for the evaluation of political measures, e.g. the obtained results suggest that strategies based on herd immunity are too risky.
\end{abstract}
\section{Introduction}
The pandemic of COVID-19 has scourged the world since the beginning of 2020\footnote{https://www.who.int/emergencies/diseases/novel-coronavirus-2019}. The responsible virus is the SARS-CoV-2, identified in China at the end of 2019 \cite{Zhou}. Governments are constantly looking for for ways to predict and contain the spread of such an illness in order to monitor the public health and to prevent economic and social issues.

Epidemic models constitute a branch of interest in applied mathematics since several years. They are adopted not only to study epidemics properly but also to predict social phenomenon or the behavior of biological systems. The simplest epidemic model is called SIR model. It looks at a population split out into three compartments: susceptible, infected and removed. The SIR model was introduced the first time in 1927 \cite{Kermack1927} and many variations have been proposed to study diseases with complex behaviors and other phenomena \cite{Murray}. Epidemic models can  also be formulated by means of the theory of stochastic processes. The first application of stochastic processes to epidemics was presented in 1955 \cite{Whittle1955} and more recently several applications have been proposed \cite{Capasso}. In some cases there is an equivalence between the two approaches \cite{Allen}. A novel model based on an operatorial approach as in quantum mechanics can be found in \cite{Bagarello}.

Concerning the new pandemic, many mathematical models have been proposed. In \cite{Ansumali2020,Calafiore2020,Giordano2020} some deterministic epidemic models for COVID-19 based on ordinary differential equations have been proposed. In \cite{Zang2020} a stochastic dynamic model has been introduced. Finally, in \cite{Faranda2020,Rihan2020} authors propose mathematical models based on stochastic differential equations.

In this paper we would like to introduce some epidemic models based on stochastic processes, taking into account peculiarities of the COVID-19 disease. We proposed two models. In the first one we consider that COVID-19 has an incubation period in which people are apparently healthy and after that they become infected and are also able to infect other people. Since COVID-19 has a quite high fatality rate, the removed people have been split in two sub-classes: healed and dead. We suppose that an individual recovers or dies after a fixed time from infection. In the second model we would like to include asymptomatic people, i.e. infectious individuals without severe or identifiable symptoms. They seem to play an important role in the diffusion of the virus because usually they don't know to be infectious. Since it is not clear whether during the incubation an individual is infectious or not, in this model we make the assumption that it is possible. Moreover, it is not ascertained so far  whether and for how long people preserve the immunity to the virus. Therefore, we consider the possibility for a healed individual to lose immunity and becomes susceptible again.

The plan of the paper is as follows. In Sec. \ref{SEC:SIRD} and \ref{SEC:SAILRD} the two stochastic models are introduced; in Sec. \ref{SEC:MC} we present the Monte Carlo algorithm adopted for simulations; in Sec. \ref{SEC:Deterministic} we propose a deterministic model to assess the validity of the one introduced in \ref{SEC:SIRD}; in Sec. \ref{SEC:Num_res} we show and comment the numerical results.

\section{A SIRD model for COVID-19 disease}
\label{SEC:SIRD}
Let us consider a fixed (no births and no deaths) population of $N$ individuals split out into four compartments:  susceptible $S$,  infected $I$, removed $R$ and  dead people $D$. In principle, it is also possible to include the so-called vital dynamics by introducing  birth and death rates but in the typical time scale of the pandemic spread the effects can be considered negligible. We suppose that the number of individuals in each class evolves in time $t\in[0,+\infty[$ because of two mechanisms: susceptible individuals become infected and infected individuals recover or die. We call the introduced model SIRD. 

To describe the infection mechanism, we suppose that the rate of new infectious cases is proportional to the number of  susceptible individuals  $S(t)$ times the fraction of infected people $I(t)/N$. The proportionality factor of  is denoted  $\beta>0$ which represents the average number of contacts of a person per unit time (the day in our case). 

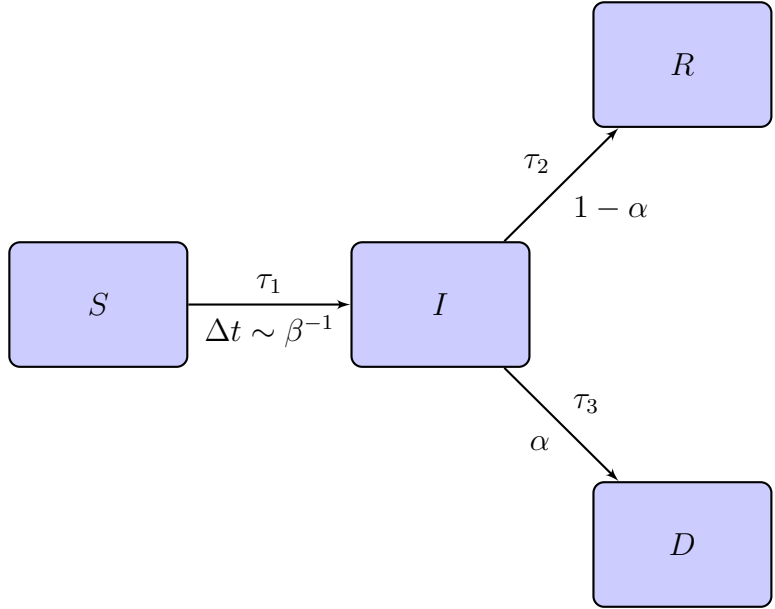
\begin{figure}[ht]
\centering
\begin{tikzpicture}[node distance=2cm, auto, thick]
\node [block] (S) {$S$};
\node [block, right of=S, node distance=4.5cm] (I) {$I$};
\node [block, above right of=I, node distance=4.5cm] (R) {$R$};
\node [block, below right of=I, node distance=4.5cm] (D) {$D$};
\path [line] (S) -- (I) node[midway,below] {$\Delta t \sim \beta^{-1}$} node[midway,above] {$\tau_1$};
\path [line] (I) -- (R) node[midway,below right] {$1-\alpha$} node[midway,above left] {$\tau_2$};
\path [line] (I) -- (D) node[midway,below left] {$\alpha$} node[midway,above right] {$\tau_3$};
\end{tikzpicture}
\caption{Flow chart of the SIRD model. Above the arrows between the compartment $I$ and the compartments $R$, $D$ we indicate the times for the transition while below the arrows the probabilities of changing compartment. $\tau_1$ is the delay time for getting infected and $\Delta t \sim \beta^{-1}$ is the frequency of the contacts between the individuals of $S$ and $I$.  }\label{FIG:SIRD}
\end{figure}

In relation to the recovery mechanism, we suppose that an infected individual has a probability $\alpha\in[0,1]$ to die and $1-\alpha$ to heal. The situation is schematized in Figure~\ref{FIG:SIRD}.

Moreover, let $\tau_1$ be the incubation time which, in this model, is the time after that an individual who contracted the virus is infected and becomes infectious. Finally, let $\tau_2$ and $\tau_3$ be the heal and dead time respectively, counted after the incubation period. First, we will assume that 
$\tau_1,  \tau_2, \tau_3$ are constant; in a second step this assumption will be relaxed to simulated effects as those related to the lockdown. 

Let us consider  $\left( S,I,R,D \right)$ as four random variables which take values in $\mathbb{N}_0^4$ and depend on time $t$. The epidemic is triggered by a small number of infected individuals $I_0$, that is at time $t=0$ we have
\begin{equation*}
(S,I,R,D)=(N-I_0,I_0,0,0).
\end{equation*}

We split the set $(S,I,R,D)$ into three subsets in each of which there will be defined a stochastic process. The first set is $\left( S,I \right)$, corresponding to the infection mechanism. We think of the two random variables not synchronized in time. In particular, we set
\begin{align*}
\tilde{I}(t) & = I(t+\tau_1)
\end{align*}
and we consider the new set $\left( S,\tilde{I} \right)$ and suppose that at time $t$ the random variables take values $\left( s,i \right)$, that is
\begin{equation}
\left( S(t),\tilde{I}(t) \right) = \left( s,i \right).
\end{equation}
After a small period of time $\Delta t>0$, $\Delta t \ll \displaystyle{\min_{i=1, \cdots,3}} \tau_i$ the state of system changes in
\begin{equation}
\left( S(t+\Delta t),\tilde{I}(t+\Delta t) \right) = \left( s+m,i+n \right),
\end{equation}
being $m,n\in\left\lbrace -1, 0, +1 \right\rbrace$.

The transition probability is defined as follows
\begin{align*}
&p_{(s,i)\to(s+m,i+n)}(\Delta t) = P\Big( \left( S(t+\Delta t),\tilde{I}(t+\Delta t) \right) = \left( s+m,i+n \right) \, \Big\vert \, \left( S(t),\tilde{I}(t) \right) = \left( s,i \right) \Big).
\end{align*}
In this way we define a continuous time Markov chain and the transition probability can be written as
\begin{equation*}
p_{(s,i)\to(s+m,i+n)}(\Delta t) = \left\lbrace
\begin{alignedat}{2}
&\beta s \frac{i}{N} \Delta t + o(\Delta t), && \qquad(m,n)=(-1,+1)\\
&1-\left( \beta s \frac{i}{N} \right)\Delta t + o(\Delta t), && \qquad(m,n)=(0,0)\\
&o(\Delta t), && \qquad\mbox{otherwise}.
\end{alignedat}
\right.
\end{equation*}

The second set of random variables is $\left( I,R \right)$, coupled to the first one. We set
\begin{align*}
\tilde{\tilde{I}}(t) & = I(t+\tau_1+\tau_2),\\
\tilde{\tilde{R}}(t) & = R(t+\tau_1+\tau_2)
\end{align*}
and consider the couple $\left( \tilde{\tilde{I}},\tilde{\tilde{R}} \right)$. Let us suppose that at time $t$ the random variables $\left( \tilde{\tilde{I}},\tilde{\tilde{R}} \right)$ take values $(i,r)$, that is
\begin{equation}
\left( \tilde{\tilde{I}}(t),\tilde{\tilde{R}}(t) \right) = (i,r).
\end{equation}
After a small period of time $\Delta t>0$ the state of the system changes in
\begin{equation}
\left( \tilde{\tilde{I}}(t+\Delta t),\tilde{\tilde{R}}(t+\Delta t) \right) = (i+n,r+u),
\end{equation}
being $n,u\in\left\lbrace -1, 0, +1 \right\rbrace$. Moreover, since a healing at time $t+\tau_1+\tau_2$ is related to an infection in the past interval $[t+\tau_1,t+\tau_1+\Delta t]$, we need to know the values assumed by the random variable $\tilde{I}$ at $t$ and $t+\Delta t$. In this case the transition probability is defined as follows
\begin{align*}
p_{(i,r)\to(i+n,r+u)}(\Delta t) = P\Big( & \left( \tilde{\tilde{I}}(t+\Delta t),\tilde{\tilde{R}}(t+\Delta t) \right) = \left( i+n,r+u \right) \, \Big\vert \, \left( \tilde{\tilde{I}}(t),\tilde{\tilde{R}}(t) \right) = \left( i,r \right),\\
& \tilde{I}(t)=j, \tilde{I}(t+\Delta t)=j+1 \Big).
\end{align*}
Therefore, a non-Markovian continuous time stochastic process is defined with the transition probability
\begin{equation*}
p_{(i,r)\to(i+n,r+u)}(\Delta t) = \left\lbrace
\begin{alignedat}{2}
&(1-\alpha) \Delta t + o(\Delta t), && \qquad(n,u)=(-1,+1)\\
&1-(1-\alpha) \Delta t + o(\Delta t), && \qquad(n,u)=(0,0)\\
&o(\Delta t), && \qquad\mbox{otherwise}.
\end{alignedat}
\right.
\end{equation*}

Finally, the third set of random variables is $\left( I,D \right)$, which is also coupled to the first one. Now we set
\begin{align*}
\hat{I}(t) & = I(t+\tau_1+\tau_3),\\
\hat{D}(t) & = D(t+\tau_1+\tau_3)
\end{align*}
and consider the couple $\left( \hat{I}, \hat{D} \right)$. Let us suppose that at time $t$ the random variables take values $(i,d)$, that is
\begin{equation}
\left( \hat{I}(t), \hat{D}(t) \right) = (i,d).
\end{equation}
After a small period of time $\Delta t>0$ the state of the system changes in
\begin{equation}
\left( \hat{I}(t+\Delta t), \hat{D}(t+\Delta t) \right) = (i+n,d+v),
\end{equation}
being $n,v\in\left\lbrace -1, 0, +1 \right\rbrace$. Moreover, since a death at time $t+\tau_1+\tau_3$ is related to an infection in the past interval $[t+\tau_1,t+\tau_1+\Delta t]$, we need to know the values assumed by the random variable $\tilde{I}$ at $t$ and $t+\Delta t$. In this case the transition probability is defined as follows
\begin{align*}
p_{(i,d)\to(i+n,d+v)}(\Delta t) = P\Big( & \left( \hat{I}(t+\Delta t), \hat{D}(t+\Delta t) \right) = \left( i+n,d+v \right) \, \Big\vert \, \left(\hat{I}(t), \hat{D}(t) \right) = \left( i,d \right),\\
& \tilde{I}(t)=j, \tilde{I}(t+\Delta t)=j+1 \Big).
\end{align*}
Therefore, a non-Markovian continuous time stochastic process is defined with the transition probability
\begin{equation*}
p_{(i,d)\to(i+n,d+v)}(\Delta t) = \left\lbrace
\begin{alignedat}{2}
&\alpha \Delta t + o(\Delta t), && \qquad(n,v)=(-1,+1)\\
&1-\alpha \Delta t + o(\Delta t), && \qquad(n,v)=(0,0)\\
&o(\Delta t), && \qquad\mbox{otherwise}.
\end{alignedat}
\right.
\end{equation*}

\section{A SAI(L)RD model for COVID-19}
\label{SEC:SAILRD}
At variance with the SIRD, the second model we are going to introduce also contemplates another two compartments: the class $A$ of asymptomatic individuals and the class $L$ of isolated infected people. We call it SAI(L)RD model. Its detailed features are summarized below
\begin{itemize}
\item[A1.]
We suppose that if an effective contact occurs between an infected (symptomatic or not) and a susceptible individual then the latter becomes infected and infectious at the same time. 
\item[A2.]
We suppose there exists a probability $\eta\in[0,1]$  to be asymptomatic and, consequently, $1-\eta$ is the probability to show symptoms. Moreover, we assume that, after a certain time $\tau_1$, a symptomatic individual is recognized and isolated into a subclass, called $L$, of lonely individuals of $I$.
\item[A3.] 
Lonely individuals are not infectious anymore. Asymptomatic people heal after a time $\tau_2$. Lonely symptomatic individuals can die with probability $\alpha$ in a time $\tau_4$ or heal with probability $1-\alpha$ in a time $\tau_3$. 
\item[A4.] Finally, removed individuals become immune to COVID-19 for a short period of time or forever. Let $\lambda\in[0,1]$ be the probability that COVID-19 confers a short immunity of time length $\tau_5$, after which recovered individuals come back to the class of susceptible people and, in principle,  can suffer a reinfection.
\end{itemize} 

\begin{figure}[ht]
\centering
\begin{tikzpicture}[node distance=2cm, auto, thick]
\node [block] (S) {$S$};
\node [block, above right of=S, node distance=4.5cm] (A) {$A$};
\node [block, below right of=S, node distance=4.5cm] (I) {$I$};
\node [block, right of=I, node distance=4.5cm] (L) {$L$};
\node [block, above right of=L, node distance=4.5cm] (R) {$R$};
\node [block, below right of=L, node distance=4.5cm] (D) {$D$};
\path [line] (S) -- (A) node[midway,below right] {$\eta$} node[midway,above left] {$\Delta t\sim\beta^{-1}$};
\path [line] (A) -- (R) node[midway,above] {$\tau_2$};
\path [line] (S) -- (I) node[midway,above right] {$\Delta t\sim\beta^{-1}$} node[midway,below left] {$1-\eta$};
\path [line] (I) -- (L) node[midway,above right] {$\tau_1$};
\path [line] (L) -- (R) node[midway,below right] {$1-\alpha$} node[midway,above left] {$\tau_3$};
\path [line] (L) -- (D) node[midway,below left] {$\alpha$} node[midway,above right] {$\tau_4$};
\path [line] (R) -- (S) node[midway,below] {$\lambda$} node[midway,above] {$\tau_5$};
\end{tikzpicture}
\caption{Flow chart of the SAI(L)RD model. Above the arrows we indicate the times for the transition while below the arrows the probabilities of changing compartment.}\label{FIG:SAILRD}
\end{figure}
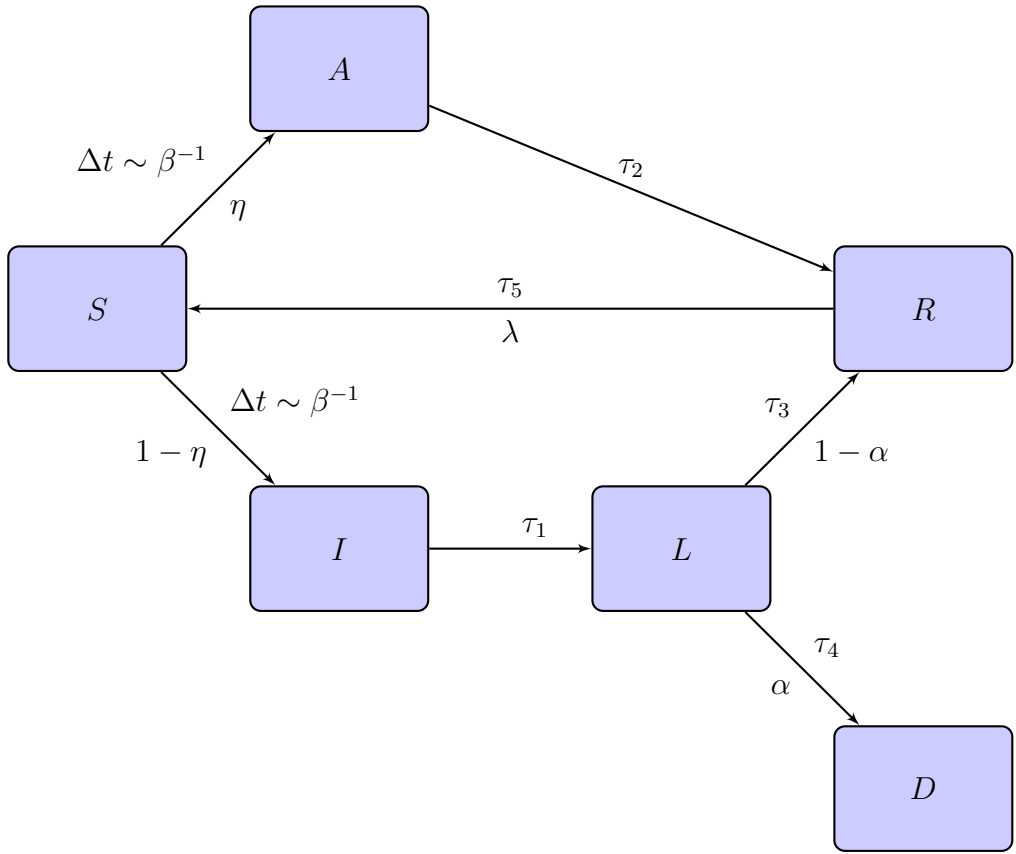

The situation is represented in Figure~\ref{FIG:SAILRD}.

We consider $\left( S,A,I,L,R,D \right)$ as a six dimensional random variable which can assume values in $\mathbb{N}_0^6$ and depends on time $t\geq0$. Now we split the set of random variables into several coupled sub-systems.
\begin{itemize}
\item
The first set is $(S,A,I)$ corresponding to the infection mechanism. We suppose that at time $t$ the random variables take values $(s,a,i)$, that is
\begin{equation}
\Big( S(t),A(t),I(t) \Big) = (s,a,i).
\end{equation}
After a small period of time $\Delta t>0$, $\Delta t \ll \displaystyle{\min_{i=1, \cdots,5}} \tau_i$, the state of the system changes in
\begin{equation}
\Big( S(t+\Delta t),A(t+\Delta t),I(t+\Delta t) \Big) = (s+m,a+k,i+n),
\end{equation}
with $m,k,n\in\left\lbrace -1,0,1 \right\rbrace$.

The transition probability is given by
\begin{align*}
p_{(s,a,i)\to(s+m,a+k,i+n)}(\Delta t) =  P\Big( & \left( S(t+\Delta t),A(t+\Delta t),I(t+\Delta t) \right) = \left( s+m,a+k,i+n \right) \\
&\, \Big\vert \, \left( S(t),A(t),I(t) \right) = \left( s,a,i \right) \Big).
\end{align*}
In this way we define a time continuous  Markov chain with transition probabilities which can be written as
\begin{equation*}
p_{(s,a,i)\to(s+m,a+k,i+n)}(\Delta t) = \left\lbrace
\begin{alignedat}{2}
&\eta\beta s \frac{i}{N} \Delta t + o(\Delta t), && \quad \mbox{if} \quad(m,k,n)=(-1,+1,0)\\
&(1-\eta)\beta s \frac{i}{N} \Delta t + o(\Delta t), && \quad \mbox{if} \quad(m,k,n)=(-1,0,+1)\\
&1-\left( \beta s \frac{i}{N} \right)\Delta t + o(\Delta t), && \quad \mbox{if} \quad(m,k,n)=(0,0,0)\\
&o(\Delta t), && \quad\mbox{otherwise}.
\end{alignedat}
\right.
\end{equation*}

\item
In our model, we suppose that all the infected individuals will be isolated after a time $\tau_1$. Let us consider $(I,L)$. We set
\begin{align*}
&\tilde{I}(t) = I(t+\tau_1),\\
&\tilde{L}(t) = L(t+\tau_1).
\end{align*}
In this case a lone infected individual at time $t + \Delta t + \tau_1$ is related to a symptomatic infection in the past interval $[t,t+\Delta t]$. Regarding the transition probability 
\begin{align*}
p_{(i,l)\to(i+n,l+j)}(\Delta t) =  P\Big( & \left( \tilde{I}(t+\Delta t),\tilde{L}(t+\Delta t) \right) = \left( i+n,l+j \right) \\
&\, \Big\vert \, \left( \tilde{I}(t),\tilde{L}(t) \right) = \left( i,l \right), I(t)=f,I(t+\Delta t)=f+1 \Big)
\end{align*}
for some nonnegative integer $f$, 
we have
\begin{equation*}
p_{(i,l)\to(i+n,l+j)}(\Delta t) = \left\lbrace
\begin{alignedat}{2}
&1 ,  && \quad \mbox{if} \quad(n,j)=(-1,+1)\\
&0, && \quad\mbox{otherwise}.
\end{alignedat}
\right.
\end{equation*}

\item
Now we consider the pair of random variables $(L,R)$. To define the healing process we set
\begin{align*}
\tilde{\tilde{L}}(t) = L(t+\tau_1+\tau_3),\\
\tilde{\tilde{R}}(t) = R(t+\tau_1+\tau_3)
\end{align*}
and suppose that 
\begin{align*}
\Big( \tilde{\tilde{L}}(t),\tilde{\tilde{R}}(t) \Big) = (l,r).
\end{align*}
After a small period of time $\Delta t >0$ the state of the system changes in
\begin{align*}
\Big( \tilde{\tilde{L}}(t+\Delta t),\tilde{\tilde{R}}(t+\Delta t) \Big) = (l+j,r+u)
\end{align*}
with $j,u\in\left\lbrace -1,0,1 \right\rbrace$. Moreover, since in this case a healing at time $t + \Delta t + \tau_1+\tau_3$ is related to an infection with symptoms in the past interval $[t,t+\Delta t]$, we need to know the values assumed by the random variable $I$ at $t$ and $t+\Delta t$. This is also equivalent to knowing the values assumed by the random variable $L$ at $t+\tau_1$ and $t + \Delta t + \tau_1$. In this case the transition probability is defined as
\begin{align*}
p_{(l,r)\to(l+j,r+u)}(\Delta t) =  P\Big( & \left( \tilde{\tilde{L}}(t+\Delta t), \tilde{\tilde{R}}(t+\Delta t) \right) = \left( l+j,r+u \right) \\
&\, \Big\vert \, \left(\tilde{\tilde{L}}(t),\tilde{\tilde{R}}(t) \right) = \left( l,r \right), L(t+\tau_1)=f,L(t + \Delta t + \tau_1) = f+1 \Big),
\end{align*}
for some nonnegative integer $f$.
In this way a time continuous  non-Markovian stochastic process is defined whose  transition probability can be written as
\begin{equation*}
p_{(l,r)\to(l+j,r+u)}(\Delta t) = \left\lbrace
\begin{alignedat}{2}
&(1-\alpha)\Delta t + o(\Delta t),  && \quad \mbox{if} \quad(j,u)=(-1,+1)\\
&1-(1-\alpha)\Delta t + o(\Delta t),  && \quad \mbox{if} \quad(j,u)=(0,0)\\
&o(\Delta t), && \quad\mbox{otherwise}.
\end{alignedat}
\right.
\end{equation*}

\item
With the same arguments, a dead process is described by $(\hat{L},\hat{D})$ where
\begin{align*}
\hat{L}(t) = L(t+\tau_1+\tau_4),\\
\hat{D}(t) = D(t+\tau_1+\tau_4).
\end{align*}
In this case a death at time $t+ \Delta t + \tau_1+\tau_4$ is related to an infection with symptoms in the past interval $[t,t+\Delta t]$ and thus the gain of one unit to the variable $L$ in $[t+\tau_1,t+\Delta t + \tau_1]$. The transition probability 
\begin{align*}
p_{(l,d)\to(l+j,d+v)}(\Delta t) =  P\Big( & \left( \hat{L}(t+\Delta t),\hat{D}(t+\Delta t) \right) = \left( l+j,d+v \right) \\
&\, \Big\vert \, \left( \hat{L}(t),\hat{D}(t) \right) = \left( l,d \right), L(t+\tau_1)=f,L(t+\Delta t +\tau_1)=f+1 \Big),
\end{align*}
for some nonnegative integer $f$,
is given by
\begin{equation*}
p_{(l,d)\to(l+j,d+v)}(\Delta t) = \left\lbrace
\begin{alignedat}{2}
&\alpha\Delta t + o(\Delta t),  && \quad \mbox{if} \quad(j,v)=(-1,+1)\\
&1-\alpha\Delta t + o(\Delta t),  && \quad \mbox{if} \quad(j,v)=(0,0)\\
&o(\Delta t), && \quad\mbox{otherwise}.
\end{alignedat}
\right.
\end{equation*}

\item
A further process we introduce is the healing of an asymptomatic individual. Let us consider  $(A,R)$. We set
\begin{align*}
&\hat{A}(t) = A(t+\tau_2),\\
&\hat{R}(t) = R(t+\tau_2).
\end{align*}
In this case a healing at time $t+\tau_2 + \Delta t$ is related to an asymptomatic infection in the past interval $[t,t+\Delta t]$. The transition probability\begin{align*}
p_{(a,r)\to(a+k,r+u)}(\Delta t) =  P\Big( & \left( \hat{A}(t+\Delta t),\hat{R}(t+\Delta t) \right) = \left( a+k,r+u \right) \\
&\, \Big\vert \, \left( \hat{A}(t),\hat{R}(t) \right) = \left( a,r \right), A(t)=f,A(t+\Delta t)=f+1 \Big),
\end{align*}
for some nonnegative integer $f$,
reads
\begin{equation*}
p_{(a,r)\to(a+k,r+u)}(\Delta t) = \left\lbrace
\begin{alignedat}{2}
&1,  && \quad \mbox{if} \quad(k,u)=(-1,+1)\\
&0, && \quad\mbox{otherwise}.
\end{alignedat}
\right.
\end{equation*}

\item
The last process we are going to introduce is the one involving a removed individual who comes back to the class of susceptible people after a certain time. To define the process, we set
\begin{align*}
&\hat{\hat{S}}(t)=S(t+\tau_5),\\
&\hat{\hat{R}}(t)=R(t+\tau_5)
\end{align*}
Let us suppose that at time $t$ the two-dimensional random variable $(\hat{\hat{S}},\hat{\hat{T}})$ takes the value $(s,r)$.
After a small period of time $\Delta t>0$ the state of the system changes in
\begin{equation*}
(\hat{\hat{S}}(t+\Delta t),\hat{\hat{R}}(t+\Delta t))=(s+m,r+u),
\end{equation*}
with $m,u\in\left\lbrace -1,0,1 \right\rbrace$. Moreover, since in this case a healed individual can come back to the class of susceptible people after a certain time $\tau_5$, it is needed to know the values assumed by the random variable $R$ at time $t$ and $t+\Delta t$. In this case the transition probability 
\begin{align*}
p_{(s,r)\to(s+m,r+u)}(\Delta t) =  P\Big( & \left(\hat{\hat{S}}(t+\Delta t),\hat{\hat{R}}(t+\Delta t) \right) = \left( s+m,r+u \right) \\
&\, \Big\vert \, \left( \hat{\hat{S}}(t),\hat{\hat{R}}(t) \right) = \left( s,r \right), R(t)=f,R(t+\Delta t)=f+1 \Big),
\end{align*}
for some nonnegative integer $f$,
is given by
\begin{equation*}
p_{(s,r)\to(s+m,r+u)}(\Delta t) = \left\lbrace
\begin{alignedat}{2}
&\lambda + o(\Delta t),  && \quad \mbox{if} \quad(m,u)=(+1,-1)\\
&1-\lambda + o(\Delta t),  && \quad \mbox{if} \quad(m,u)=(0,0)\\
&o(\Delta t), && \quad\mbox{otherwise}.
\end{alignedat}
\right.
\end{equation*}
\end{itemize}

The major advantage to adopt a stochastic model is the possibility to easily add further more sophisticated features. Indeed, the delays are assumed constants but it is possible to consider in turn  the times $\tau_i$ as random variables obeying suitable probability distributions. However, in average we get the same results. 

\section{The Monte Carlo method for stochastic process simulation}
An efficient simulation of both the  SIRD and SAI(L)RD models can be performed by a Monte Carlo approach. The details are outlined in the next subsections.

\label{SEC:MC}
\subsection{SIRD model}
Firstly we describe the method adopted for SIRD model. The state of the system is represented by a time-dependent random vector variable
\begin{equation}
\mathbf{X}(t) = \left(X_1 (t), X_2 (t), \cdots, X_N(t) \right)\in D^N,
\end{equation}
for $t\geq0$, where $D=\left\lbrace -1, 0, 1, 2 \right\rbrace$ and $N$ is the population size. We indicate by $X_i(t)\in D$ for $i=1,\ldots,N$ the trajectory of an individual in time, i.e. the time-evolution of the states assumed by the $i$-th person. $D$ is a set of labels where 0 represents a susceptible individual, 1 an infected one, 2 a healed person and $-1$ a dead individual. 

At time $t=0$ a number of $I_0$ infected individuals are labeled by 1 randomly, all the others are susceptible thus labeled by 0. 
We note that the pure process of the encounters is Markovian, no matter it leads to an infection or not. Therefore, for each infected individual a contact time $t$ is determined according to the exponential distribution of scale parameter $\beta$, that is
\begin{equation}\label{EQ:Exp_dt}
 t = -\frac{1}{\beta}\log \xi,
\end{equation}
 $\xi$ being a random number uniformly distributed in $[0,1]$. 
 
 Let us suppose that the minimum contact time is that of the $j$-th individual, $ t_{j,1}$ (the second index indicates the first temporal step of the individual $j$).
 At this point  another individual $i$ is chosen randomly. If it belongs to the susceptible class then the contact is effective and after the incubation, i.e. a period of time $\tau_1$, the individual $i$ changes its state in infectious, 
\begin{equation}
X_i(t_{j,1}+\tau_1)=1.
\end{equation}
At time $t_{j,1} +\tau_1$, the destiny $d\in\left\lbrace -1,2\right\rbrace$ of the new infectious is established too accordingly to a Bernoulli distribution with probability $\alpha$, that is $d\sim\mathcal{B}(1,\alpha)$. If the destiny is to heal then after a time $\tau_2$ the individual state changes from infectious to recovered; if the destiny is to die then after a time $\tau_3$ the state of the $i$-th individual changes from infectious to dead:
\begin{equation}
\begin{alignedat}{2}
& X_i(t_{j,1} +\tau_1+\tau_2)=2 && \qquad\mbox{if}\quad d=2,\\
& X_i(t_{j,1} +\tau_1+\tau_2)=-1 && \qquad\mbox{if}\quad d=-1.
\end{alignedat}
\end{equation}

After the choice of the individual $i$, the individual $j$ still continues to infect unless in the meantime he has recovered or passed away. Another random infection time $t_{j,2}$ is generated according to (\ref{EQ:Exp_dt})  and we set
$$t_j= t_{j,1} + t_{j,2}.$$
Once again we determine the infected individual having associated the minimum time and iterate the procedure.
 The algorithm ends whether there are no more susceptible individuals. 
 
 In order to record the time evolution of the system, a time grid is fixed and at each time of such a grid  we count the number of individuals in the several classes. Moreover, to reduce the statistical noise an averaging procedure is applied as follows. We perform the entire simulation $k$ times. Let $T_r$, $r = 1, 2, \cdots, k$, be the time at which the algorithm ends at the $r$th simulation. We set
\begin{equation}
m_k = \frac{T_1+T_2+\ldots+T_k}{k},
\end{equation}
the average of the final process times. After introducing the error as
\begin{equation}
\varepsilon_k = \vert m_{k+1}-m_k \vert,
\end{equation}
as stopping criterion we adopt
\begin{equation}
\varepsilon_k<\mbox{tol}, \qquad k>N_{min}.
\end{equation}
Here $\mbox{tol}$ is a numerical tolerance and   $N_{min}$  is a minimum number of iterations which are required  to prevent early stops of the numerical procedure.

\subsection{SAI(L)RD model}
In a similar way the SAI(L)RD model can be simulated. Now we have
\begin{equation*}
\mathbf{X}(t)\in E^N,
\end{equation*}
where $E=\left\lbrace -1,0,1,2,3,4 \right\rbrace$. The label $-1$ represents a dead person,  $0$ a susceptible individual, $1$ an infected one, $2$ a healed person, $3$ an asymptomatic individual, $4$ a lone infected one. 

Even in this case, at time $t=0$ we randomly select $I_0$ individuals we label as infected, i.e. by $1$. 
A contact time $t$ is determined according to \eqref{EQ:Exp_dt} for each infected individual. Let us suppose that the minimum contact time is that of the $j$-th individual, $t_{j,1}$. At this point  another individual $i$ is chosen randomly. Now at variance with the SIRD model,  if the latter  is susceptible her/his state changes as follows: we determine  the symptomaticity $s\in\left\lbrace 1,3\right\rbrace$  by a Bernoulli distribution having probability $\eta\in[0,1]$. If $s=3$ the individual is asymptomatic and, after the  time  $\tau_2$, she/he will heal, and therefore
\begin{equation}
X_i(t_{j,1}+\tau_2) = 2.
\end{equation}
If $s=1$ the person is symptomatic infected and, after the time $\tau_1$, she/he will be isolated, that is
\begin{equation}
X_i(t_{j,1}+\tau_1) = 4.
\end{equation}
Moreover, the destiny $d\in\left\lbrace -1,2\right\rbrace$ of such an individual is established according to a Bernoulli distribution with probability $\alpha$. If the destiny is to heal then after a time $\tau_3$ the individual state changes from infectious to recovered; if the destiny is to die then after a time $\tau_4$ the individual state changes from infectious to dead. That evolution can be described as follows
\begin{equation}
\begin{alignedat}{2}
& X_i(t_{j,1}+\tau_1+\tau_3)=2 && \qquad\mbox{if}\quad d=2,\\
& X_i(t_{j,1}+\tau_1+\tau_4)=-1 && \qquad\mbox{if}\quad d=-1.
\end{alignedat}
\end{equation}
         Finally there is also the possibility that a healed individual  loses the immunity, coming back to the susceptible class. We take that into account by generating a random number according to a Bernoulli distribution of parameter  $\lambda \in (0,1)$. Furthermore,  if the immunity is lost, one has two possibilities: \\
 if the individual suffered from an asymptomatic infection we set
\begin{equation}
X_i(t_{j,1} + \tau_2+\tau_5) = 0 
\end{equation}
 if she/he suffered from a symptomatic infection we set
\begin{equation}
X_i(t_{j,1} +\tau_1+\tau_3+\tau_5) = 0 
\end{equation}

After the choice of the individual $i$, the individual $j$ still continues to infect unless in the meantime she/he has recovered without losing the immunity or passed away. Another random infection time $t_{j,2}$ is generated according to (\ref{EQ:Exp_dt})  and we set
$$t_j= t_{j,1} + t_{j,2}.$$
Again we determine the infected individual having associated the minimum time and iterate the procedure.
The algorithm ends when all individuals can change their state no longer. 

To reduce the statistical noise we adopted the same technique as the SIRD model presented above.

\section{A deterministic SIRD model with delays}
\label{SEC:Deterministic}
In order to check the validity of the  SIRD stochastic model proposed above, a deterministic delayed SIRD model is devised as well. Since the disease has an incubation time $\tau_1$, the number of susceptible people decreases by a quantity depending on the amount of infected at a previous time $t-\tau_1$; the amount of recovered and dead people after a time $\tau_2$ and $\tau_3$ respectively is proportional to the amount of people who have been infected at the previous time $t-\tau_2$ and $t-\tau_3$ respectively. From those considerations, we propose the following model 
\begin{subequations}
\begin{align}[left = \empheqlbrace\,]
\dot{S}(t)=&-\beta S(t)\frac{I(t-\tau_1)}{N}H(t-\tau_1)\\
\dot{I}(t)=&\beta S(t)\frac{I(t-\tau_1)}{N}H(t-\tau_1)+(1-\alpha)\dot{S}(t-\tau_2)+\alpha\dot{S}(t-\tau_3)\\
\dot{R}(t)=&-(1-\alpha)\dot{S}(t-\tau_2)\\
\dot{D}(t)=&-\alpha\dot{S}(t-\tau_3)
\end{align}
\end{subequations}
where $H(\cdot)$ represents the Heaviside step function. In that way we include the effects of an incubation time and healing or death times. 

For further analysis it is convenient to work with proportions. After the substitution
$$
S \mapsto \frac{S}{N}, I \mapsto \frac{I}{N}, R \mapsto \frac{R}{N}, D \mapsto \frac{D}{N}
$$ 
and some simple algebraic  manipulations,  the system can be  written more explicitly as
\begin{subequations} \label{system}
\begin{align}[left = \empheqlbrace\,]
\dot{S}(t)=&-\beta S(t) I(t-\tau_1) H(t-\tau_1)\\
\dot{I}(t)=&\beta S(t) I(t-\tau_1) H(t-\tau_1) - (1-\alpha)\beta S(t-\tau_2) I(t-\tau_1-\tau_2) H(t-\tau_1-\tau_2)\\
&-\alpha\beta S(t-\tau_3)I(t-\tau_1-\tau_3) H(t-\tau_1-\tau_3)\nonumber\\
\dot{R}(t)=&(1-\alpha)\beta S(t-\tau_2) I(t-\tau_1-\tau_2) H(t-\tau_1-\tau_2)\\
\dot{D}(t)=&\alpha\beta S(t-\tau_3) I(t-\tau_1-\tau_3) H(t-\tau_1-\tau_3)
\end{align}
\end{subequations}

The system must be augmented assigning the functions
\begin{equation*}
\begin{alignedat}{2}
S(t)&=\Phi_1(t) && \qquad t\in[- \max (\tau_2,\tau3),0],\\
I(t)&=\Phi_2(t) && \qquad t\in[- \max (\tau_1+ \tau_2, \tau_1 + \tau3),0].
\end{alignedat}
\end{equation*}
As customary we assume that $\Phi_1$ and $\Phi_2$ are continuous in the  considered intervals.
Regarding the other variables  it is realistic to take $R(0) = D(0) = 0$. Specifically we assume 
$\Phi_1 (t) = S_0 \in ]0,1[$ and  $\Phi_2 (t) = I_0 = 1 - S_0 \in ]0,1[$ with $I_0 \ll 1$. Therefore at $t=0$ we have
\begin{equation*}
S(0) + I(0) + R(0) + D(0) = 1.
\end{equation*}

The presence of the delays makes the qualitative analysis of the system rather cumbersome, so a complete phase portrait is a daunting task. However, some  insights can be deduced anyway. 

Along the solution of the system (\ref{system})
$$
S(t) + I(t) + R(t) + D(t) = 1.
$$
In fact, summing up the equations (\ref{system}a)-(\ref{system}d) one has
$$
\dot Y(t)  = 0,
$$
where $Y=S+I+R+D$. Since $Y(0) =1$, it follows that $Y(t) = 1$ $\forall t > 0$. 

The equilibrium points of the system (\ref{system}) are:\\
\begin{itemize}
\item 
Endemic solutions: 
\begin{equation}
S=0, I = I^*, R = R^*, D = D^*. 
\end{equation}
with $I^*,R^*,D^* \in [0,1]$ satisfying $I^*+ R^*+ D^* = 1$.
\item 
Disease-free solutions: 
\begin{equation}
S=S^*, I = 0, R = R^*, D = D^*. 
\end{equation}
with $S^*,R^*,D^* \in [0,1]$ satisfying $S^*+ R^*+ D^* = 1$.
\end{itemize}

If we linearize around the generic endemic critical point, one gets the following characteristic equation for the eigenvalues
$$
\lambda \left(\lambda + \beta I^*  \right) =0
$$
which shows that the endemic stationary states are linearly stable for any $I^* > 0$.  

The linearization around the disease-free stationary points leads to a much more complex characteristic equation
$$
\lambda \left[\lambda -  \beta S^* \left(e^{- \tau_1 \lambda} + (1 - \alpha) e^{- (\tau_1 + \tau_2)  \lambda} + \alpha e^{- (\tau_1 + \tau_3)  \lambda} \right) \right] =0;
$$
which in general admits infinite solutions in the complex plane due to the functional nature of the equations. The only viable way to get the eigenvalues is to resort to a numerical procedure \cite{libro_proceedings}. 
Therefore, since the primary goal is to have a comparison with the stochastic model, we look directly at the numerical solutions of the system (\ref{system}).
To this aim, we adopt a first order finite differences scheme. 

Let us fix a temporal grid $0=t_0<t_1<\ldots<t_M=T_{max}$ of constant time step $\Delta t$. We introduce the numerical approximations 
\begin{align*}
S_k \approx S(t_k),\quad
I_k \approx I(t_k),\quad
R_k \approx R(t_k),\quad
D_k \approx D(t_k),
\end{align*}
for $k=0,1,\ldots,M$, and  discretize the system (\ref{system}) as follows
\begin{equation*}
\begin{aligned}
S_{k+1} = & S_k - \Delta t \beta S_k \frac{I_{k_1}}{N}H(t_k-\tau_1),\\
I_{k+1} = & I_k + \Delta t \left\lbrace \beta S_k \frac{I_{k_1}}{N}H(t_k-\tau_1) - (1-\alpha)\beta S_{k_2}\frac{I_{k_{12}}}{N}H(t_k-\tau_1-\tau_2)\right.\\
&\left. - \alpha\beta S_{k_3}\frac{I_{k_{13}}}{N}H(t_k-\tau_1-\tau_3)\right\rbrace,\\
R_{k+1} = & R_k + \Delta t (1-\alpha)\beta S_{k_2}\frac{I_{k_{12}}}{N}H(t_k-\tau_1-\tau_2),\\
D_{k+1} = & D_k + \Delta t \alpha\beta S_{k_3}\frac{I_{k_{13}}}{N}H(t_k-\tau_1-\tau_3).
\end{aligned}
\end{equation*}
where  the indexes $k_1$, $k_2$, $k_{12}$, $k_3$ and $k_{13}$ are given by
\begin{equation*}
\begin{alignedat}{2}
k_j & = \max\left\lbrace 0, \left\lfloor \frac{t_k-\tau_j}{\Delta t} \right\rfloor \right\rbrace, && \qquad j=1,2,3,\\
k_{1m} & = \max\left\lbrace 0, \left\lfloor \frac{t_k-\tau_1-\tau_m}{\Delta t} \right\rfloor \right\rbrace, && \qquad m=2,3,
\end{alignedat}
\end{equation*}
with $\lfloor \cdot \rfloor$ the floor function.

Note that at each time step the condition
$$
S_{k+1} + I_{k+1} + R_{k+1} + D_{k+1} =  S_{k} + I_{k} + R_{k} + D_{k} 
$$
is satisfied.

\section{Numerical results}
\label{SEC:Num_res}
\subsection{SIRD model}
Concerning the SIRD model  introduced in Sec. \ref{SEC:SIRD}, we perform some numerical simulations by adopting the algorithm of Sec. \ref{SEC:MC}. A crucial point is to fix the parameters entering the model. About the mortality $\alpha$ we consider the infection fatality ratio (IFR), i.e. the ratio between the number of deaths from disease and the number of infected individuals, whose value is reported in the range 0.5 - 1\% \cite{WHO_mortality}. According to \cite{WHO_incubation}, the incubation period $\tau_1$ is on average 5-6 days, but it can be as long as 14 days. For the healing and dead time $\tau_2$ and $\tau_3$ respectively, we remark that the commonly adopted criteria for discharging patients from isolation are the following: for symptomatic patients, 10 days after the symptom onset, plus at least 3 additional days without symptoms; for asymptomatic individuals, 10 days after positive test \cite{WHO_times}. In all the simulations of the present paper we have assumed $I_0=1$. The adopted values are reported in Table~\ref{TAB:SIRD}.
\begin{table}[ht]
\centering
\begin{tabular}{cc}
\hline
\textbf{Parameter} & \textbf{Value}\\
\hline
$\alpha$ & 0.006\\
$\tau_1$ & 6 d\\
$\tau_2$ & 12 d\\
$\tau_3$ & 13 d\\
\hline
\end{tabular}
\caption{Parameters adopted for the simulation of the SIRD model. The $\tau$'s are in days (d).}
\label{TAB:SIRD}
\end{table}

\begin{figure}[ht]
\centering
\includegraphics[width=0.49\textwidth]{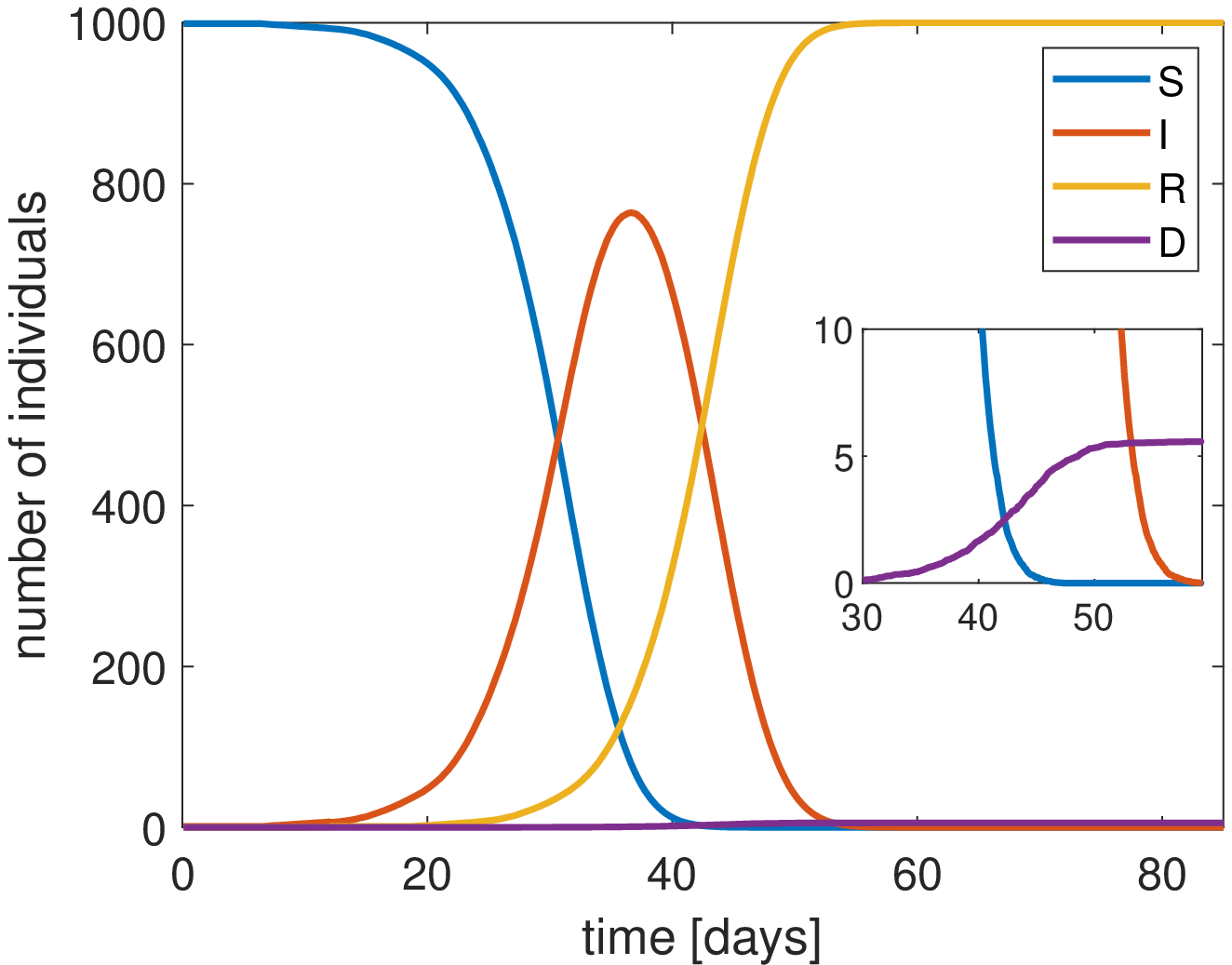}
\includegraphics[width=0.49\textwidth]{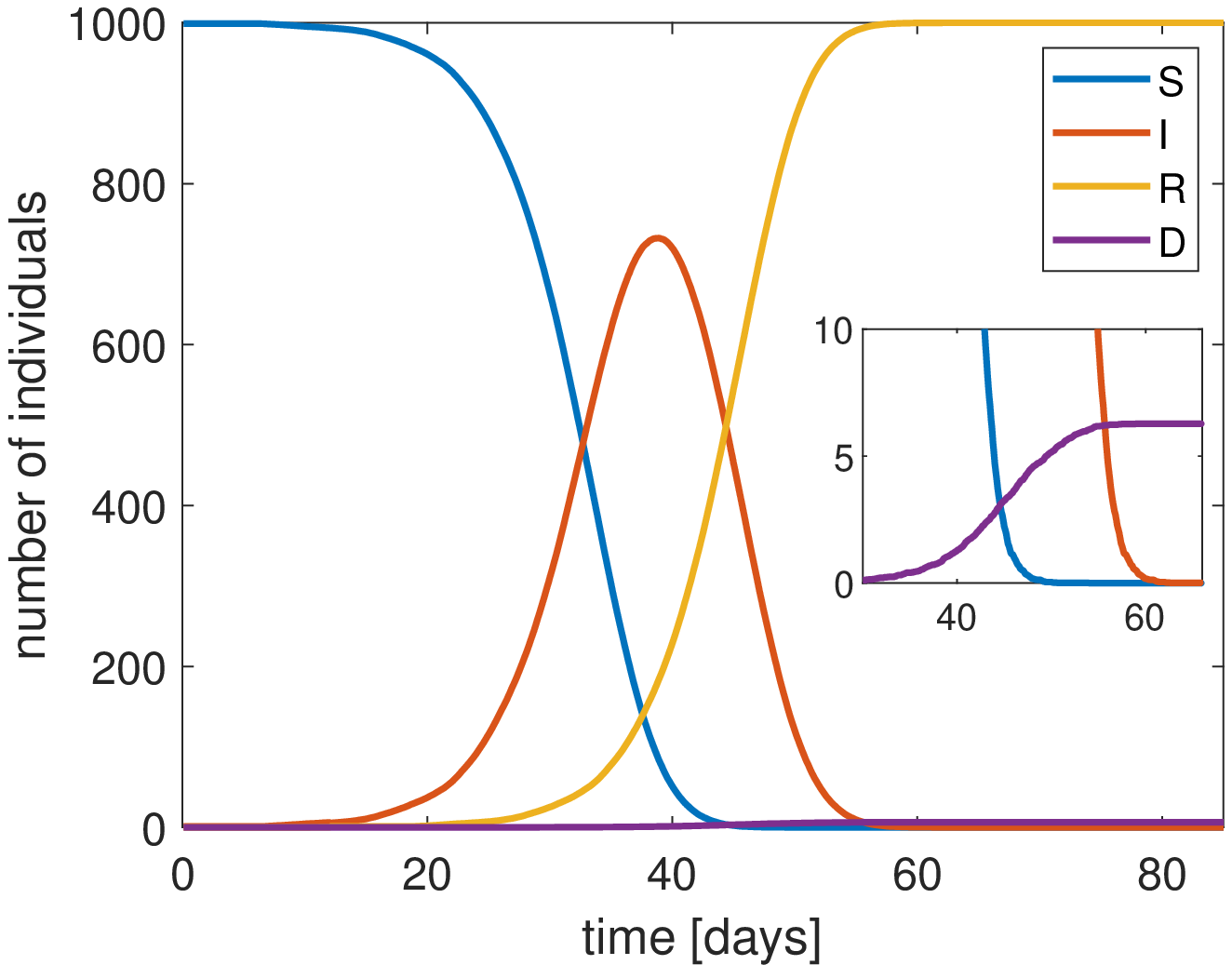}\\
\includegraphics[width=0.49\textwidth]{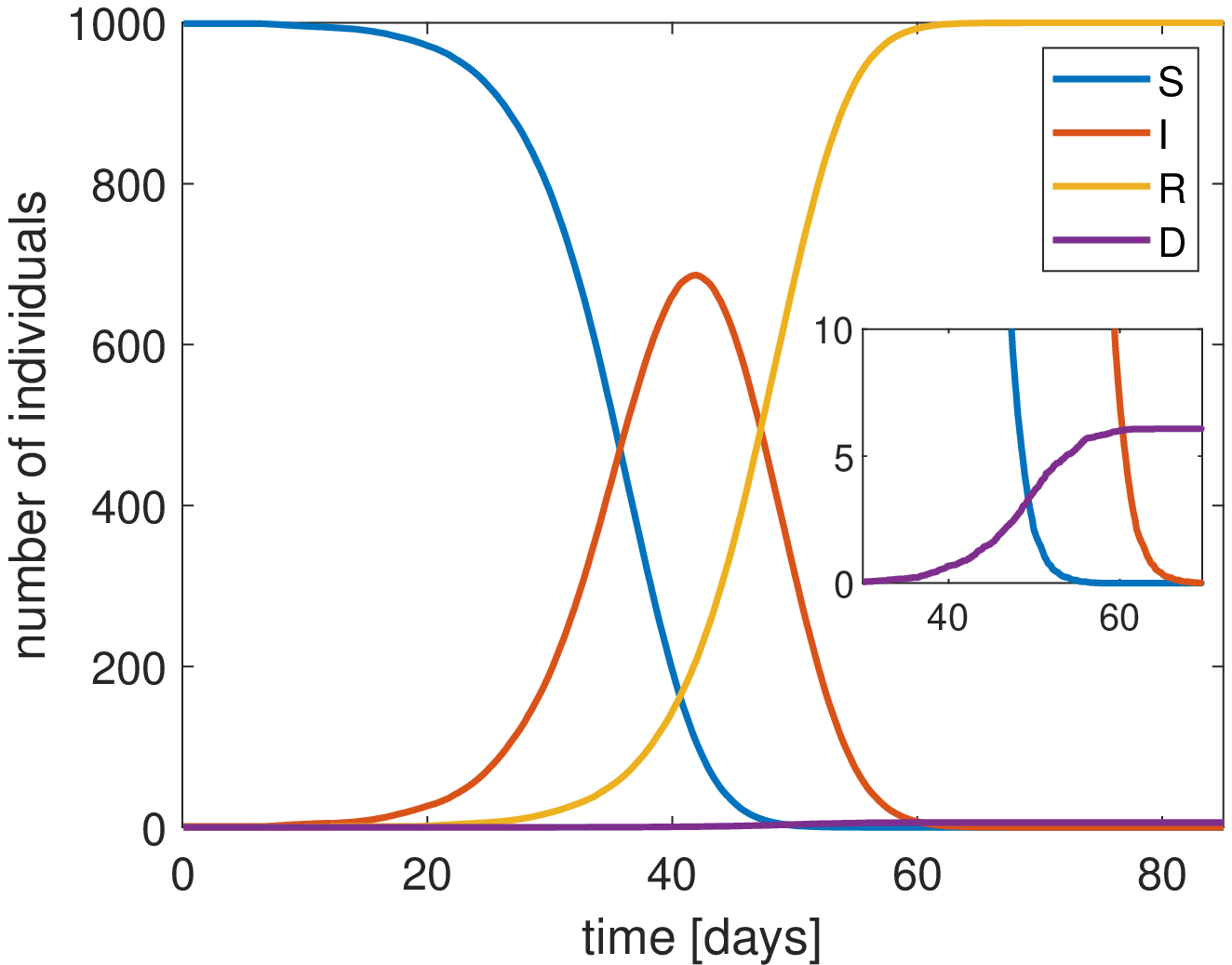}
\includegraphics[width=0.49\textwidth]{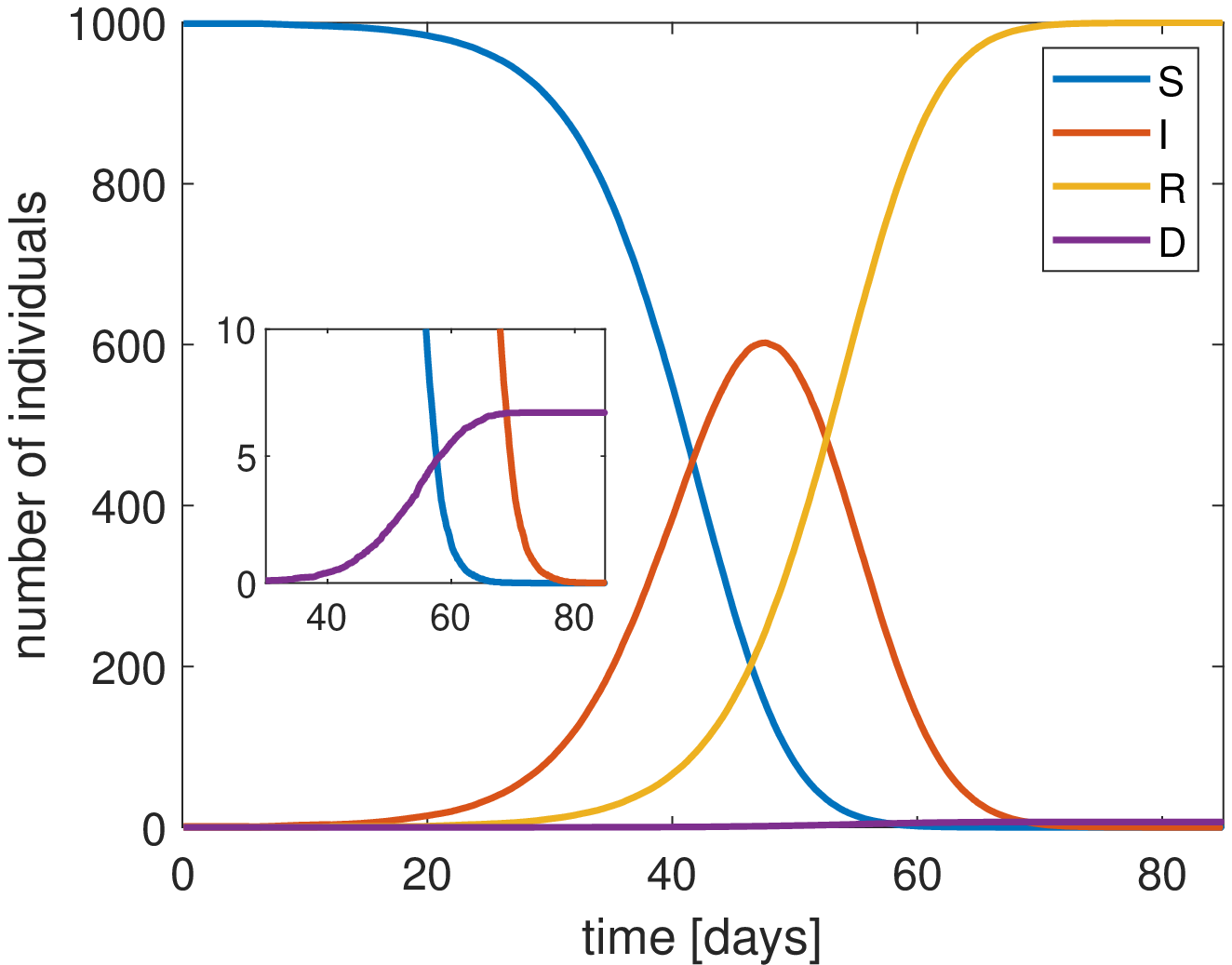}
\caption{Behavior of SIRD model with a population of 1000 individuals and $\beta$ of 1.2 (top-left), 1 (top-right), 0.8 (bottom-left) and 0.6 (bottom-right). In the inset the values for large times are magnified. 
}
\label{FIG:SIRD_beta}
\end{figure}

More controversial is  to fix the contact frequency $\beta$. It should be around  $\beta\simeq 1$ d$^{-1}$ according to some estimations \cite{Peng2020} but it varies with time and as consequence of measures of social restriction by the authorities. For such a reason we have performed the simulations of the stochastic SIRD model for several values of this parameter: $\beta=1.2,1,0.8,0.6$. The results are shown in Fig. \ref{FIG:SIRD_beta} in the case of a population of 1000 individuals. The same cases have been also simulated when $N=10000$. The qualitative behavior is essentially the same, in particular the value of the maximum percentage of infected, but with a temporal dilation. 
\begin{figure}[ht]
\centering
\includegraphics[width=0.49\textwidth]{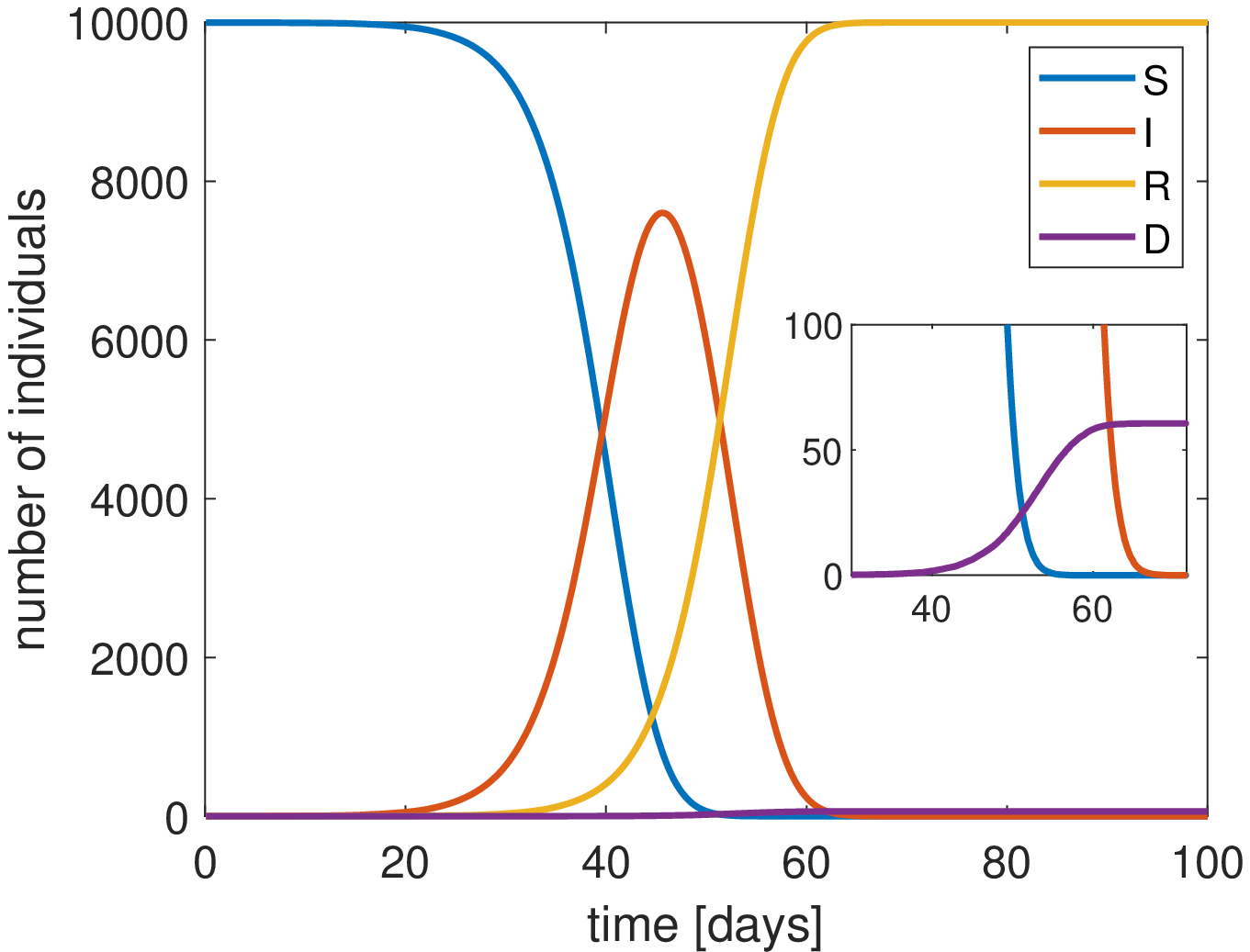}
\includegraphics[width=0.49\textwidth]{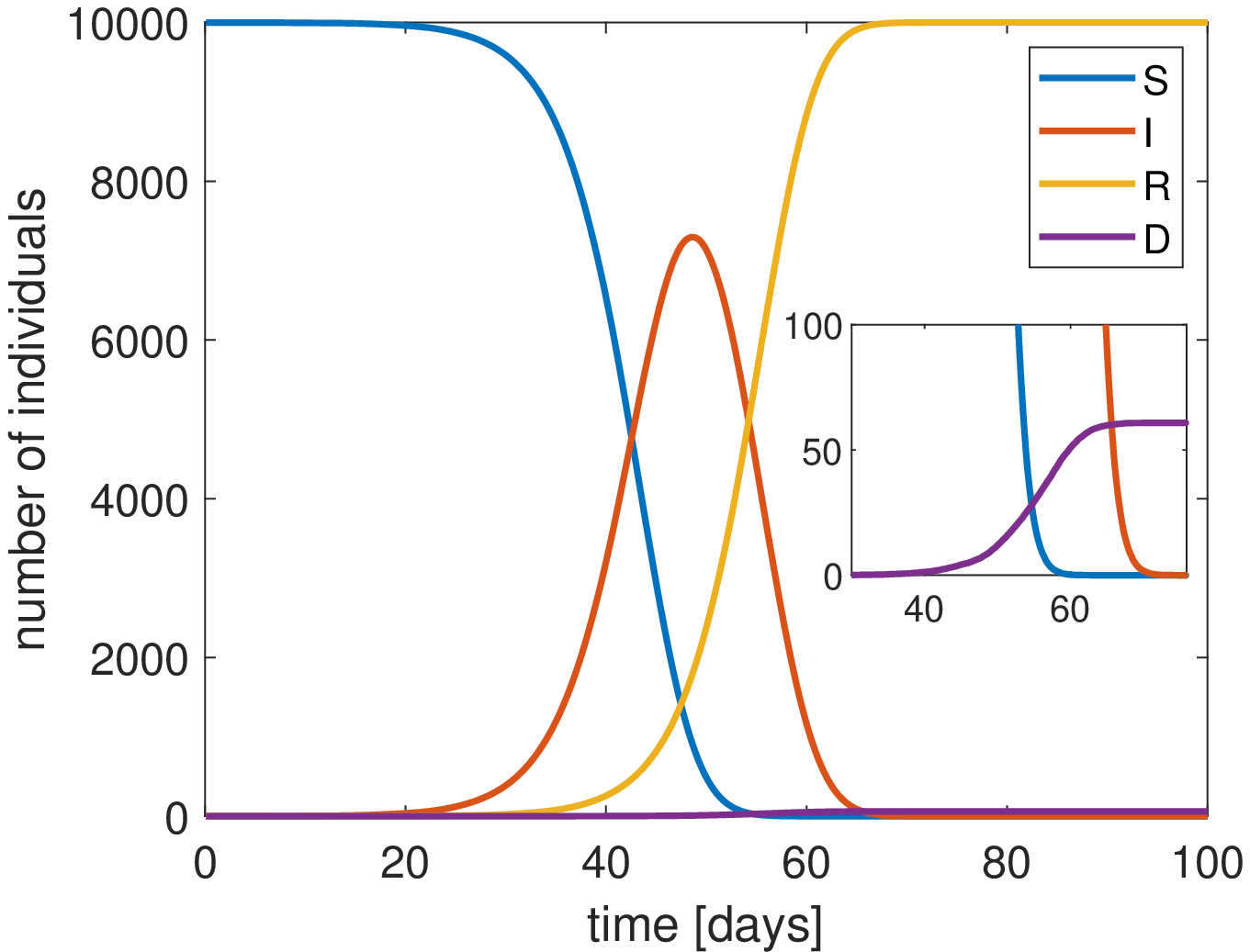}\\
\includegraphics[width=0.49\textwidth]{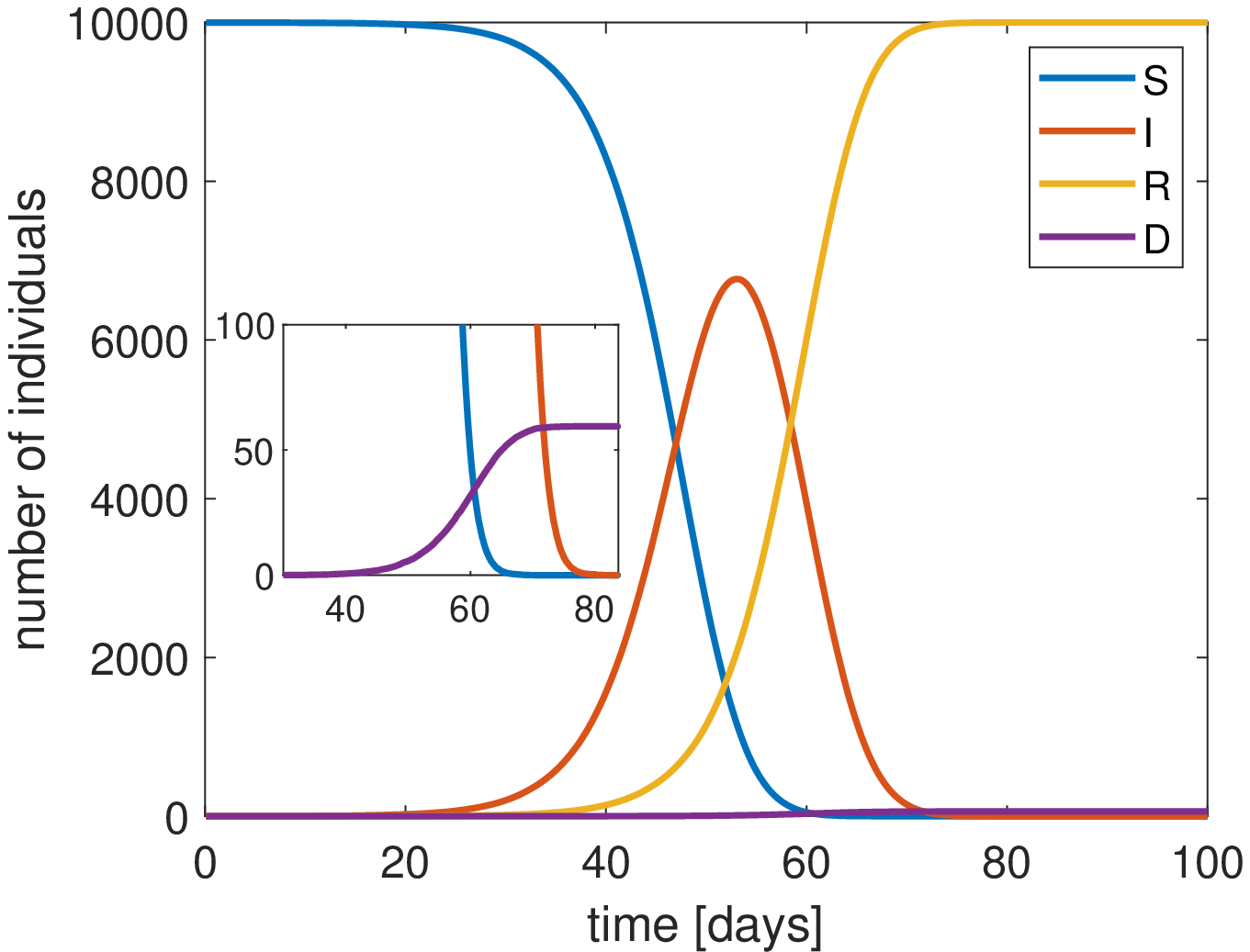}
\includegraphics[width=0.49\textwidth]{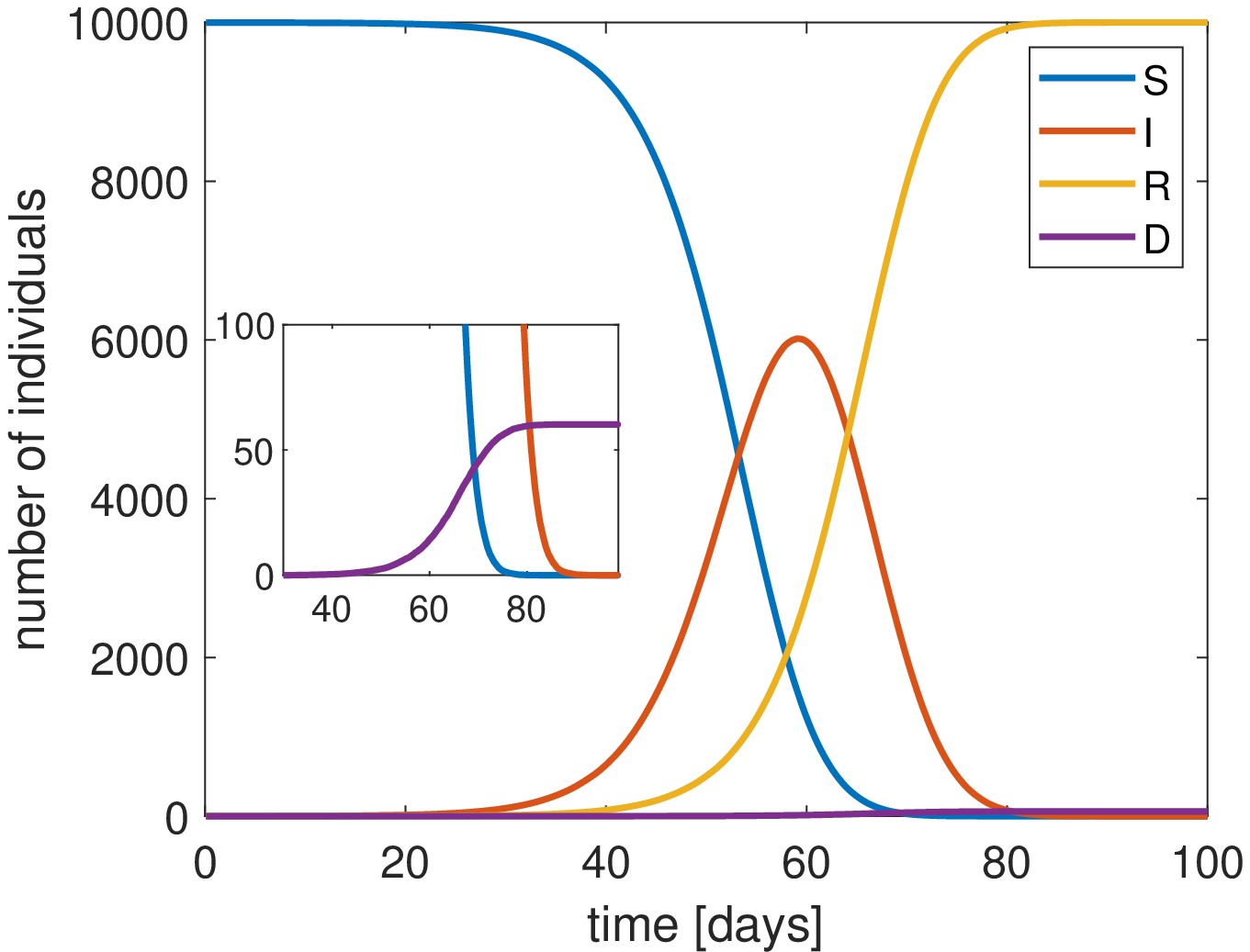}
\caption{Behavior of SIRD model with a population of 10000 individuals and $\beta$ of 1.2 (top-left), 1 (top-right), 0.8 (bottom-left) and 0.6 (bottom-right). In the inset the values for large times are magnified. 
}
\label{FIG:SIRD_beta_10k}
\end{figure}

Typical features of all the simulations are the reduction of susceptible people and a non monotone behaviour of the number of infected ones. The latter first increases and then tends to zero. The peak of the infected people is  considered (see \cite{Fine}) as the value which represents the state when the {\em herd immunity} is reached  without any actions by the authorities in charge for the health issues, also according to other models, e.g. see \cite{Ansumali2020}.  From a quantitative point of view our results indicate that the herd immunity is reached when about  70$\div$80 \% of the population is infected which is a quite pessimistic foresight. Therefore, political strategies based on a pure herd immunity appear too risky because they could lead to a 0.6\% of dead people due to the COVID-19 (see the inset in Fig. \ref{FIG:SIRD_beta}). 

\begin{figure}[ht]
\centering
\includegraphics[width=0.49\textwidth]{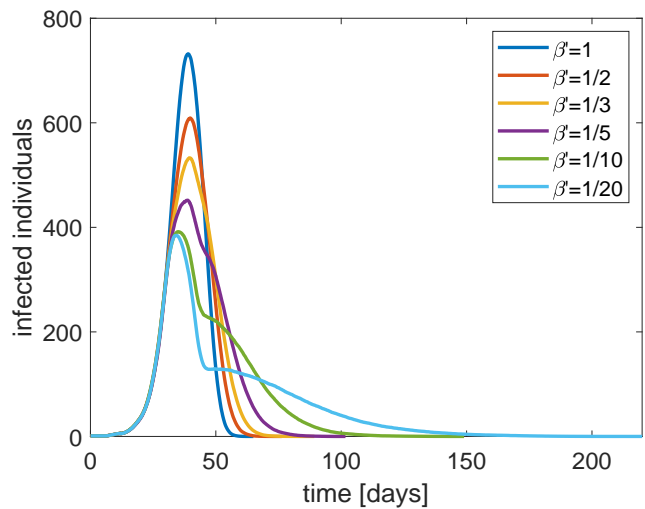}
\caption{Behavior of the SIRD model with a population of 1000 individuals and several values of $\beta'$: number of infected individuals versus days. Note that the peak lowers by decreasing $\beta'$.}
\label{FIG:SIRD_lock}
\end{figure}

To avoid the drawbacks mentioned above, worldwide governments are assuming restrictions on free movement of people, the  so-called lockdown, to contain the spread of the pandemic. In order to simulate the effect of a lockdown in our stochastic SIRD model, we suppose that the parameter $\beta$ changes in $\beta'$ whether the fraction of infected individuals reaches 10\% of the total population size. Remember that $1/\beta$ is the average contact time. So, if social distancing measures are adopted, they can be modelled as a reduction of $\beta$, that is by extending the average contact time among the individuals. In Fig. \ref{FIG:SIRD_lock} we show the curve of infected individuals for several values of $\beta'$. It is evident that the peak of infected persons lowers even if we have longer tails. This is quite realistic because the disease still remains but the number of recovered people increases in a slower way. Of course, if the aim is to alleviate the burden of hospitalized patients, the presence of a longer time to get the disappearance of the disease is a minor matter.  Apparently with the lockdown it seems that the  herd immunity is reached with a lower percentage of infected people than the case without lockdown. 

\begin{figure}[ht]
\centering
\includegraphics[width=0.49\textwidth]{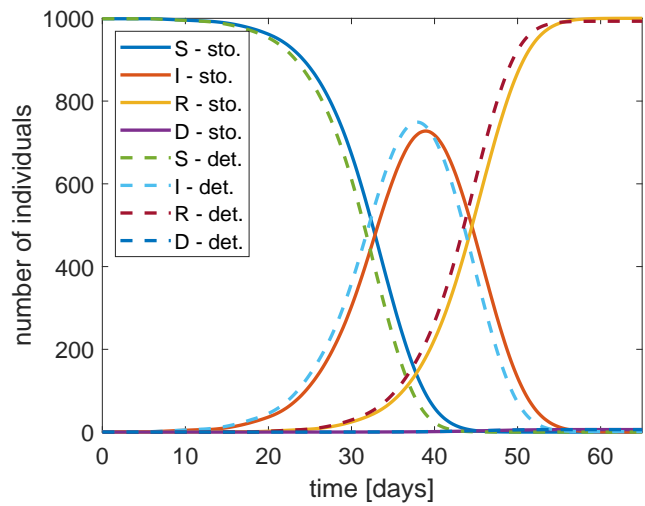}
\caption{Comparison between the stochastic (continuous lines) and deterministic (dashed lines) SIRD model in the case of $N=1000$ and $\beta=1$. Similar results are obtained with other values of $\beta$.}
\label{FIG:SIRD_comp}
\end{figure}

To assess the validity of the model, in Fig. \ref{FIG:SIRD_comp} we show a comparison between numerical results obtained by the stochastic SIRD model of Sec. \ref{SEC:SIRD} and the deterministic one presented in Sec. \ref{SEC:Deterministic}. There is a good agreement by obtaining a cross validation of both models. It is noteworthy that the numerical solution tends to the disease-free stationary critical points $S^* = I^* =0$, $R^*, D^* \in [0,1]$ with $R^* +  D^* = 1$.

\FloatBarrier

\subsection{SAI(L)RD model}
As further improvements, here we also include the presence of asymptomatic and infected isolated individuals by presenting the results of the simulations obtained with  the SAI(L)RD model. Regarding the parameters for the asymptomatic infections,  $\eta$ and $\tau_2$, the literature reports that the proportion of people who become infected and remain asymptomatic throughout infection seems to be in the range 40-45\% and they can transmit the virus for a period of  about 14 days \cite{Oran2020}. Concerning the parameter $\tau_1$, we assume that symptomatic individuals can transmit the virus  and they are isolated after an average  period of 5-6 days because the illness is detected by tests. During the isolation they can heal or die with the same arguments of the SIRD model. 

Since it is not known how long antibody responses will be maintained or whether they will provide protection from reinfection \cite{Seow2020}, we suppose that healed individuals may get a temporary or permanent immunity. We set the probability $\lambda$ to have temporary immunity equal to 0.1 and the duration  90 days. The list of the adopted values is reported in Table~\ref{TAB:SAILRD}. In the plots  $L$ is included in $I$.
\begin{table}[ht]
\centering
\begin{tabular}{cc}
\hline
\textbf{Parameter} & \textbf{Value}\\
\hline
$\alpha$ & 0.006\\
$\eta$ & 0.4\\
$\lambda$ & 0.1\\
$\tau_1$ & 6 d\\
$\tau_2$ & 14 d\\
$\tau_3$ & 12 d\\
$\tau_4$ & 13 d\\
$\tau_5$ & 90 d\\
\hline
\end{tabular}
\caption{Parameters adopted for the simulation of the SAI(L)RD model.}
\label{TAB:SAILRD}
\end{table}

\begin{figure}[ht]
\centering
\includegraphics[width=0.49\textwidth]{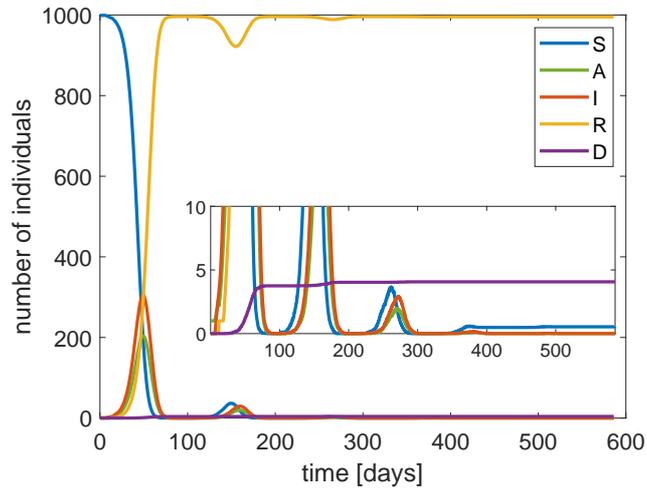}
\caption{Behavior of SAI(L)RD model with a population of 1000 individuals and $\beta$ set as 1.}
\label{FIG:SAILRD_simple}
\end{figure}

In Fig. \ref{FIG:SAILRD_simple} we show the numerical solutions of the SAI(L)RD model in the case of $N=1000$ and $\beta=1$. The main distinctive feature with respect to the results obtained by the SIRD model is that
after about 100 days we observe a new availability of susceptible individuals due to the loss of immunity. This along with a second wave of infection which, however, has a lower peak. Asymptotically we get again a disease-free situation with about 6\% of dead people.  We remark that by isolating the infected people the herd immunity is guaranteed by a peak of infected of about 
30\% to which about 20\% of asymptomatic individuals must be added with a total of about 50\% of people with disease. Again the strategy based on reaching the herd immunity can be deemed as to avoid because too costly in terms of hazard for the life of the population. 

\begin{figure}[ht]
\centering
\includegraphics[width=0.49\textwidth]{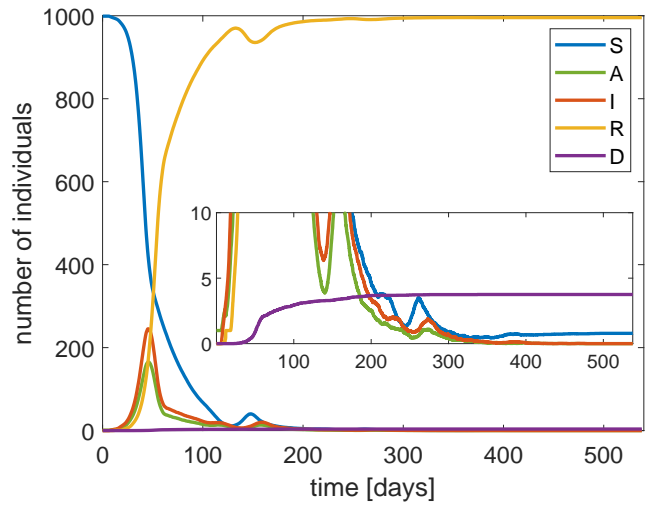}
\caption{Behavior of SAI(L)RD model with a population of 1000 individuals in the case of a lockdown.}
\label{FIG:SAILRD_lock}
\end{figure}

\begin{figure}[ht]
\centering
\includegraphics[width=0.49\textwidth]{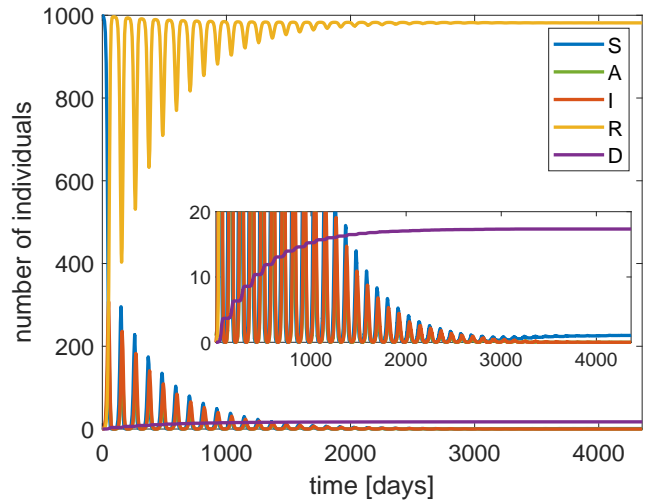}
\caption{Behavior based on the SAI(L)RD model with a population of 1000 individuals in the case of a probability of reinfection $\lambda =$ 0.8.}
\label{FIG:SAILRD_reinfections}
\end{figure}

The above findings strongly support the need of the restrictive measures from a quantitative point of view. 
In order to analyze the effect  of a lockdown we have also adopted the SAI(L)RD model. We start by taking $\beta=1$ and then we set $\beta=1/10$ when the fraction of infected individuals reaches 0.1. After 60 days we switch to $\beta=1/3$ considering some restrictions still valid after the lockdown. The obtained numerical solutions  are shown in Fig. \ref{FIG:SAILRD_lock}. 
The values of the maximum for both  infected and asymptomatic individuals are lower. Moreover,
we observe that a second wave of infection is present which is less intense than the case of constant $\beta$. However, the asymptotic number of the dead person
with the lockdown is only slightly improved. The main effect of the lockdown is to alleviate  the congestion in the intensive care because the infections are spread over a longer time.  
Finally, we would like to remark that if the probability of immunity loss $\lambda$ is very high then periodic  waves of persistent infection will show up for some years, as indicated in Fig. \ref{FIG:SAILRD_reinfections}. Note that the asymptotic number of the dead people is about 1.7\%.

As last remark, if the size of the population is greater a dilation of the time is observed but the main features remain  the same as the case of 1000 people.

%
%
%
%
%
%

%
%
\FloatBarrier
\section*{Conclusions and acknowledgments}

Two stochastic models for simulating the evolution of the pandemic SARS-CoV-2 have been proposed. By using a Monte Carlo method, realistic situations have been investigated, obtaining insights about the possibility to get the herd immunity and the effects of measures as social distancing.  

The models and the numerical approach have been tested by considering a deterministic version. The good agreement between the stochastic and deterministic results provides a cross validation.

The models are quite flexible and allow us an easy inclusion of the effects of a lockdown. The evolution we have considered does not take into account  a campaign of vaccination but this can be  included with a moderate additional effort. 

The authors G.N. and V.R. acknowledge the support from INdAM (GNFM).
%
%
%
\clearpage \noindent

\begin{thebibliography}{99}
%
\bibitem{Zhou}
P. Zhou, XL. Yang, XG. Wang, et al., \emph{A pneumonia outbreak associated with a new coronavirus of probable bat origin}, Nature, 579, 270–273 (2020).
%
\bibitem{Kermack1927}
W. O. Kermack, A. G. McKendrick, \emph{A Contribution to the Mathematical Theory of Epidemics}, Proceedings of the Royal Society of London, series A, vol. 115, no. 772 (1927).
%
\bibitem{Murray}
J. D. Murray, \emph{Mathematical Biology I. An Introduction}, Springer-Verlag New York (2002).
%
\bibitem{Whittle1955}
P. Whittle, \emph{The Outcome of a Stochastic Epidemic--A Note on Bailey's Paper}, Biometrika, vol. 42, no. 1/2, pp. 116–122 (1955).
%
\bibitem{Capasso}
V. Capasso, D. Bakstein, \emph{An Introduction to Continuous-Time Stochastic Processes. Theory, Models, and Applications to Finance, Biology, and Medicine}, Birkh\"{a}user Basel (2015).
%
\bibitem{Allen}
L. J. S. Allen, \emph{An Introduction to Stochastic Processes with Applications to Biology. Second Edition}, CRC Press - Taylor \& Francis Group (2010).
%
\bibitem{Bagarello} F. Bagarello, F. Gargano, F. Roccati, \emph{Modeling epidemics through ladder operators}, Chaos, Solitons and Fractals, 140, 110193 (2020).
%
\bibitem{Ansumali2020}
S. Ansumali, S. Kaushal, A. Kumar, M. K. Prakash, M. Vidyasagar, \emph{Modelling a pandemic with asymptomatic patients, impact of lockdown and herd immunity, with applications to SARS-CoV-2}, Annual Reviews in Control, 50, pp. 432-447 (2020).
%
\bibitem{Calafiore2020}
G. C. Calafiore, C. Novara, C. Possieri, \emph{A time-varying SIRD model for the COVID-19 contagion in Italy}, Annual Reviews in Control, 50, pp. 361–372 (2020).
%
\bibitem{Giordano2020}
G. Giordano, F. Blanchini, R. Bruno, P. Colaneri, A. Di Filippo,
A. Di Matteo, M. Colaneri, \emph{Modelling the COVID-19 epidemic and
implementation of population-wide interventions in Italy}, Nature Medicine, 26, pp. 855–860 (2020).
%
\bibitem{Zang2020}
Y. Zhang, C. You, Z. Cai, J. Sun, W. Hu, X.‑H. Zhou, \emph{Prediction of the COVID‑19 outbreak in China based on a new stochastic dynamic model}, Scientific Reports, 10, 21522 (2020).
%
\bibitem{Faranda2020}
D. Faranda, T. Alberti, \emph{Modeling the second wave of COVID-19
infections in France and Italy via a stochastic SEIR model}, Chaos 30, 111101 (2020).
%
\bibitem{Rihan2020}
F.A. Rihan, H. J. Alsakaji, C. Rajivganthi, \emph{Stochastic SIRC epidemic model with time-delay for COVID-19}, Advances in Difference Equations, 2020, 502 (2020).
%
\bibitem{WHO_mortality}
World Health Organization, Estimating mortality from COVID-19, Scientific brief, 4 August 2020, WHO Reference Number: WHO-2019-nCoV-Sci\_Brief-Mortality-2020.1
%
\bibitem{WHO_incubation}
World Health Organization, Transmission of SARS-CoV-2: implications for infection prevention precautions, Scientific brief, 9 July 2020, WHO Reference Number: WHO/2019-nCoV/Sci\_Brief/Transmission\_modes/2020.3
%
\bibitem{WHO_times}
World Health Organization, Criteria for releasing COVID-19 patients from isolation, Scientific brief, 17 June 2020, WHO Reference Number: WHO/2019-nCoV/Sci\_Brief/Discharge\_From\_Isolation/2020.1
%
\bibitem{Peng2020}
L. Peng, W. Yang, D. Zhang, C. Zhuge, and L. Hong, \emph{Epidemic analysis of COVID-19 in China by dynamical modeling}, arXiv:2002.06563 (2020).
%
\bibitem{Oran2020}
D. P. Oran, E. J. Topol, \emph{Prevalence of Asymptomatic SARS-CoV-2 Infection}, Ann Intern Med., 173, pp. 362-367 (2020).
%
\bibitem{Seow2020}
J. Seow, C. Graham, B. Merrick, et al., \emph{Longitudinal observation and decline of neutralizing antibody responses in the three months following SARS-CoV-2 infection in humans}, Nat. Microbiol., 5, pp. 1598–1607 (2020).
%
\bibitem{Fine}  P. Fine, K. Eames, D. L. Heymann,  \emph{``Herd immunity'': A rough guide}, Clinical Infectious Diseases, 52(7), 911-916 (2011).
%
\bibitem{libro_proceedings} S. Feng, M. Chen, N. Zhan, M. Fr\"anzle,
and B. Xue, \emph{Taming Delays in Dynamical Systems. Unbounded Verification of Delay Differential Equations}. In: I. Dillig, S. Tasiran (Eds.) ``Computer Aided Verification'', CAV 2019. Lecture Notes in Computer Science, vol 11561. Springer, Cham (2019).
%
\end{thebibliography}
\end{document}